\documentclass[aps,prl,twocolumn,showpacs,superscriptaddress,groupedaddress,longbibliography]{revtex4}  
\usepackage{graphicx}  
\usepackage{dcolumn}   
\usepackage{bm}        
\usepackage{amssymb}  
\usepackage{float} 
\usepackage{graphicx}
\usepackage[rgb]{xcolor}
\usepackage{amsmath}
\usepackage{pdfcomment}

\begin{document}

\title{Uncoupled dark states can inherit polaritonic properties}
\author{C. Gonzalez-Ballestero}
\affiliation{Departamento de F{\'\i}sica Te{\'o}rica de la Materia Condensada and Condensed Matter Physics Center (IFIMAC), Universidad Aut\'onoma de Madrid, E-28049 Madrid, Spain}

\author{J. Feist}
\affiliation{Departamento de F{\'\i}sica Te{\'o}rica de la Materia Condensada and Condensed Matter Physics Center (IFIMAC), Universidad Aut\'onoma de Madrid, E-28049 Madrid, Spain}

\author{E. Gonzalo Bad\'\i a}
\affiliation{Departamento de F{\'\i}sica Te{\'o}rica de la Materia Condensada and Condensed Matter Physics Center (IFIMAC), Universidad Aut\'onoma de Madrid, E-28049 Madrid, Spain}

\author{Esteban Moreno}
\affiliation{Departamento de F{\'\i}sica Te{\'o}rica de la Materia Condensada and Condensed Matter Physics Center (IFIMAC), Universidad Aut\'onoma de Madrid, E-28049 Madrid, Spain}

\author{F.J. Garcia-Vidal}
\email{fj.garcia@uam.es}
\affiliation{Departamento de F{\'\i}sica Te{\'o}rica de la Materia Condensada and Condensed Matter Physics Center (IFIMAC), Universidad Aut\'onoma de Madrid, E-28049 Madrid, Spain}
\affiliation{Donostia International Physics Center (DIPC), E-20018 Donostia/San Sebasti\'an, Spain}

\date{\today}

\begin{abstract}
When a collection of quantum emitters interacts with an electromagnetic field, the whole system can enter into the collective strong coupling regime in which hybrid light-matter states, i.e., polaritons can be created. 
Only a small portion of excitations in the emitters are coupled to the light field, and there are many dark states that, in principle, retain their pure excitonic nature. Here we theoretically demonstrate that 
these dark states can have a delocalized character, which is inherent to polaritons, despite the fact that they do not have a photonic component. This unexpected behavior only appears when the electromagnetic field displays a discrete spectrum. In this case, when the main loss mechanism in the hybrid system stems from the radiative losses of the light field, dark states are even more efficient than polaritons in transferring excitations across the structure.
\end{abstract}

\maketitle


The ability to create and engineer hybrid light-matter states, i.e., polaritons, can bring together the most advantageous properties of both worlds, such as the high speed and delocalization of photons together with the stability and interacting character of matter excitations \cite{Kimble}. In order to create such hybrid light-matter states, it is usually necessary to reach the so-called collective strong coupling (CSC) between a light field and an ensemble of quantum emitters (QEs). This CSC regime is characterized by the coupling of the electromagnetic field to a set of states in the ensemble (the bright states) forming the polaritons \cite{TormaBarnes2015}. However, many  states of the QEs stay uncoupled to the photons and are thus called dark states. Since its first experimental demonstration with Rydberg atoms \cite{HarochePRL1983}, CSC has been reached in a variety of systems, ranging from atomic beams to ion Coulomb crystals and organic materials \cite{KimblePRL1992, HemmerichPRL2003, BrenneckeNature2007,ColombeNature2007, DrewsenPRA2012, whittakerNature1998,BellessaPRL2004,EbbesenPRB2005}.  Polaritons display a wide range of basic phenomena such as superfluidity  \cite{AlbertoAmoNatPhys2009}, Bose-Einstein condensation \cite{KasprzakNature2006}, or lasing \cite{ImamogluPRA1996}. Besides fundamental prospects, polaritonic systems are also interesting for many applications that cover, among others, future quantum technologies  \cite{SanvittoNatComm2013, VuleticPRL2005,VuleticPRL2009}, both light harvesting \cite{ColesNatComm2014,GonzalezBallesteroPRBRapid2015} and transport of energy and charge in organic materials \cite{EbbesenNatMat2015,JohannesPRL2014,SchachenmayerPupillo_PRL2015},  and even control of chemical reactions \cite{Hutchison2012}.

Despite the great deal of attention received by polaritons, the uncoupled dark states have often been ignored as they are assumed not to benefit from the light-matter coupling. Indeed, these pure matter states are considered only a source of losses for polaritons \cite{delPinoNJP2015}, their potential applications being limited to passive operations such as qubit storage \cite{FleischauerLukinPRA}. In this Letter, we challenge this standard view of dark states as passive elements in the CSC regime. First, we study an extended system in which the electromagnetic (EM) spectrum is continuous. Compatible with the customary picture described above, we show that the wavefunction associated with the dark modes is strongly localized. We then analyze the case of a photonic nanostructure that supports a discrete EM spectrum where, as opposed to the previous case, the wavefunction of the dark states displays a  delocalized character, similar to that exhibited by polaritons. Moreover, we also demonstrate that if the main loss mechanism of the system resides within the EM modes, dark states can be much better excitation carriers that their polariton counterparts. 
  
In this work we consider a very general light-matter system, where we define a set of photonic modes with energies $\omega_\alpha$ and creation operators $a_\alpha^\dagger$. These modes interact with an ensemble of $N$ QEs with energies $\epsilon_j$ and fermionic operators $\sigma_j$. According to the dipole approximation, the coupling rate is proportional to both the dipole moment of the QEs and the electric field amplitude, $g_{j\alpha} = -\boldsymbol\mu_j\cdot \mathbf{E}_\alpha(\mathbf{r}_j)$. The system is described by an extension of the Jaynes-Cummings Hamiltonian \cite{carmichael} ($\hbar = 1)$,
  \begin{equation}\label{H}
  \begin{split}
  H_0 &=  \sum_{j} \epsilon_j \sigma^\dagger_j\sigma_j +\sum_{i,j} V_{ij}\left(\sigma_i^\dagger\sigma_j+\sigma_j^\dagger\sigma_i\right) \\ &+\sum_{\alpha}\omega_\alpha a^\dagger_\alpha a_\alpha + \sum_{j,\alpha}\left(g_{j\alpha}\sigma^\dagger_ja_\alpha + g_{j\alpha}^*\sigma_ja_\alpha^\dagger\right),  \end{split}
  \end{equation}
  where in the second term we include the dipole-dipole interaction between the emitters, $V_{ij}$. In order to describe the losses in both QEs and light modes, both energies $\epsilon_j$ and $\omega_\alpha$ contain a non-hermitian imaginary part \cite{carmichael}. Under driving, the full Hamiltonian of the system will be $H = H_0 + V(t)$, where $V(t)$ describes a weak coherent pump of the first QE in the ensemble, 
  \begin{equation}
  V(t) = \Omega_p\cos(\omega_L t)f(t)\left(\sigma_1^\dagger+\sigma_1\right),
  \end{equation}
  $\Omega_p \ll 1$ being the pump strength. The modulation function, $f(t)$, is assumed to vary slowly in time such that the pulse is quasi-monochromatic. We will employ the pump $V(t)$ to introduce excitations in the system in a controlled way, in order to study the steady-state properties of the wavefunction. 

For illustrative purposes, we will study two particular EM environments, namely the plasmon modes supported by an infinite silver nanowire (continuous EM spectrum) and those corresponding to a silver nanoparticle (discrete spectrum), but we stress that our findings are very general. Both metallic structures are described as cylinders of radius $r_0 = 55$ nm lying along the $z$ axis, characterized by a Drude-Lorentz permittivity $\varepsilon_m(\omega)$ \cite{suppl}, and embedded in a dielectric with $\varepsilon_d = 2.4$. The plasmon eigenmodes of the nanowire can be analytically calculated \cite{jackson1999classical}, whereas the localized surface plasmon (LSP) modes supported by the nanoparticle have been obtained numerically by using a Finite Element Method software (COMSOL Multiphysics). In order to comply with the full quantum description in Eq. (\ref{H}), the calculated plasmon modes have been adequately quantized \cite{suppl}. As for the QEs surrounding both structures,  we choose as an example similar parameters than J-aggregated molecules at room temperature \cite{MollJCHEMPHYS1995,ValleauJCHEMPHYS2012,ScwartzCPC2013}, namely energy $\text{Re}\left[\epsilon_j\right] = 1.4$ eV, dipole moment $\vert \boldsymbol{\mu}_j\vert \approx 0.75$ e $\cdot$ nm, and decay rate $\text{Im}\left[\epsilon_j\right] = -0.5 $ meV. For simplicity, we assume they are homogeneously distributed on a cylindrical layer $35$ nm above the metallic surface, with a first neighbor distance of $3$ nm, and dipole moments oriented radially.

\begin{figure}
\centering
\includegraphics[width=\linewidth]{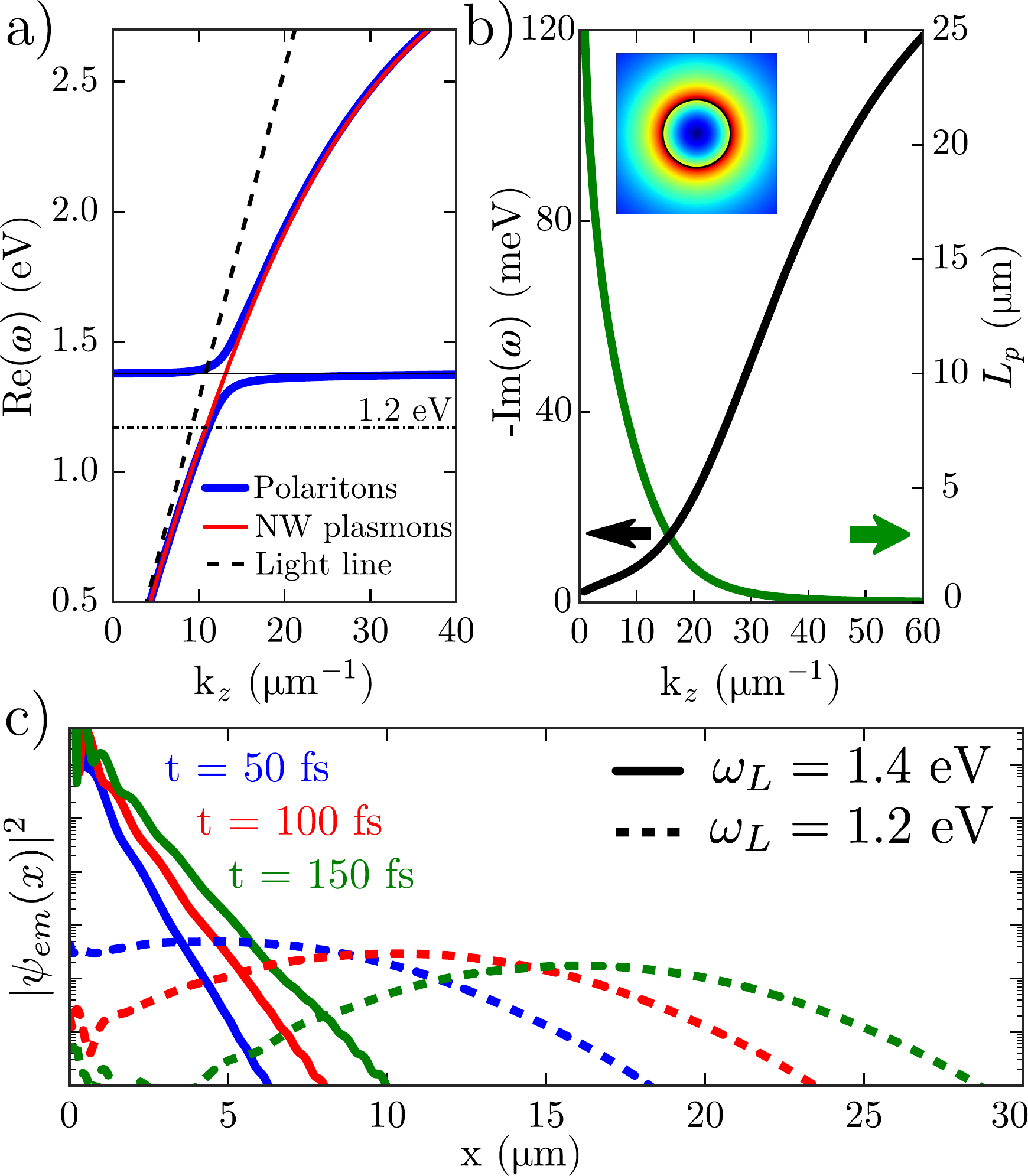}
\caption{(Color online) a) Plasmon dispersion relation of the infinite nanowire (red) and the corresponding polaritons (blue). b) Decay rate (black) and propagation length (green) of the plasmons supported by the nanowire. The inset shows the electric field norm of the fundamental mode at $1.4$ eV. c) Spatial distribution of the system wavefunction at different time intervals. Solid lines represent the diffusive behavior when pumping the flat region of the band, $\omega_L = 1.4$ eV. Dashed lines show a polariton propagating along the system ($\omega_L = 1.2$ eV).}\label{figg1}
  \end{figure}

Let us first analyze the case of an infinitely long nanowire (NW). The dispersion relation of this structure is shown in Fig. \ref{figg1}a (red line), along with the corresponding polariton dispersion (blue lines). Note that the system is in the CSC regime, since the energy separation between the two polaritons at the anticrossing point ($k_z \approx 12\;\mu$m$^{-1}$), known as Rabi splitting, is much larger than the plasmon losses Im$(\omega)$, shown in Fig. \ref{figg1}b. In this case, such losses originate from absorption in the metal, and give rise to a finite plasmonic  propagation length, displayed in the green curve in Fig. \ref{figg1}b.

For the infinite NW system, we emulate the continuum nature of the EM spectrum by imposing periodic boundary conditions over a $30\;\mu$m long unit cell, containing $N=1.88\times10^6$ QEs. We choose a finite duration pump pulse, $f(t) = e^{-(t/\tau)^2}$, where the pump is kept quasi-monochromatic through a very small frequency window, $\tau^{-1} = 0.01$ eV. Since the pumping rate $\Omega_p$ is very weak, the system wavefunction is calculated by standard perturbation theory \cite{suppl}. The solid lines in Fig. \ref{figg1}c show the spatial distribution of the QE population $\vert \psi_{em}(x,t)\vert^2$, at three different times, when the pump is tuned at the frequency of the dark states, i.e., $\omega_L = 1.4$ eV. In this case, the wavepacket is localized at the origin $x = 0$, and the probability spreads along the system in a diffusive manner due to the widening of the initial distribution. Both the strong localization and the diffusive behavior of the wavepacket are expected since the pump frequency lies on a flat region of the dispersion relation, where the group velocity is practically zero, and the associated modes have a purely excitonic, i.e., localized character. Naturally, when the pump frequency lies on a region of non-zero group velocity ($\omega_L = 1.2$ eV), a polariton propagates through the whole nanowire thanks to its photonic component, as visualized by the dashed lines in Fig. \ref{figg1}c.

 \begin{figure}
 \centering
 \includegraphics[width=\linewidth]{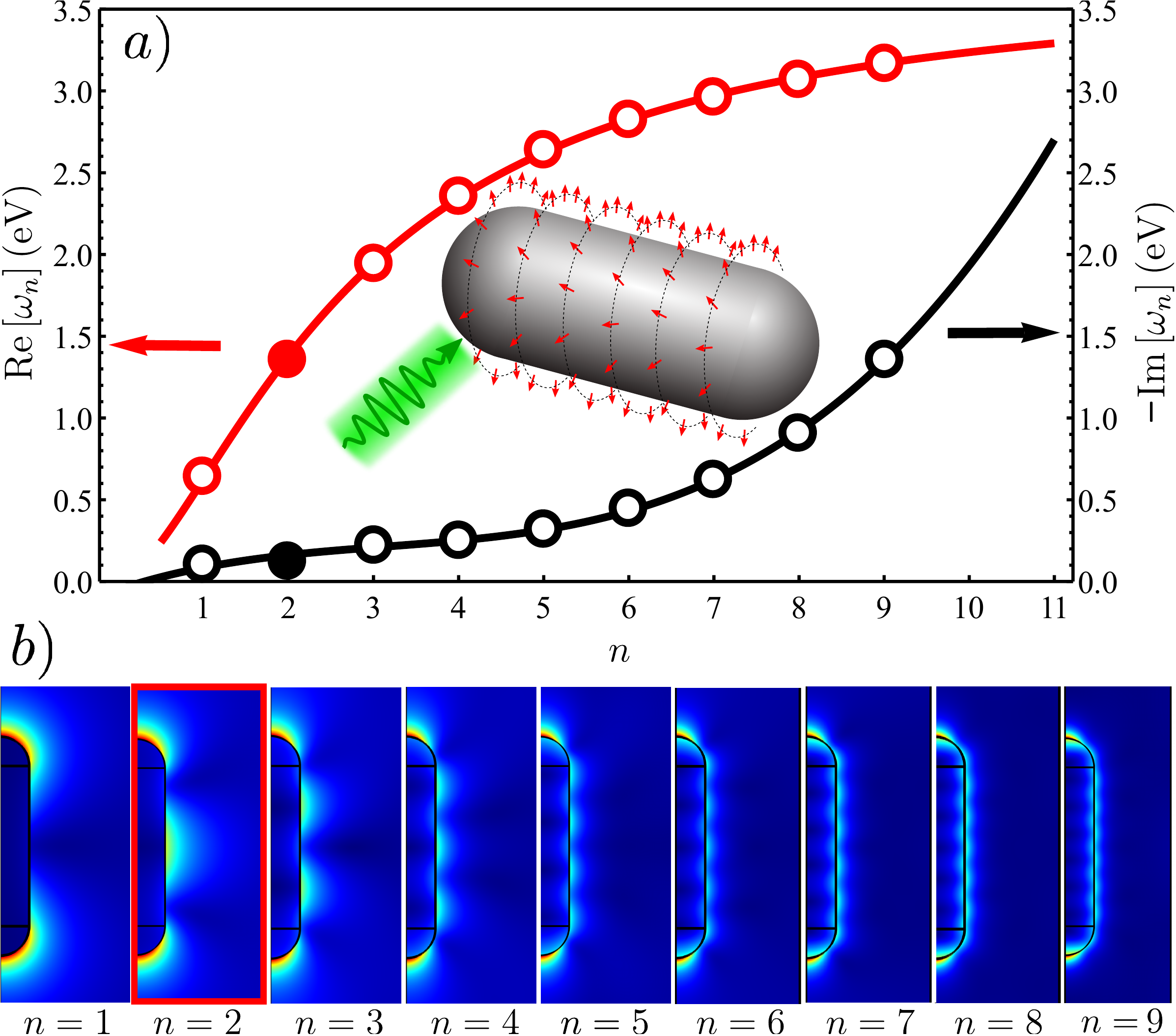}
 \caption{(Color online) a) Numerical results for the real (red) and imaginary parts (black) of the eigenmode frequencies of the metallic nanoparticle. b) Norm of the electric field for the corresponding eigenmodes, which are labeled by the mode index $n$. In our calculations we assume that the $n=2$ mode is resonant with the excitations in the QEs and this mode is highlighted in both panels.}\label{figg2}
 \end{figure}

The spatial extension of dark states, however, is very different when the ensemble of QEs interacts with a \textit{discrete} set of electromagnetic modes. In order to illustrate this, we consider a cylindrical nanoparticle (NP) $300$ nm long, terminated by two hemispherical caps as depicted schematically in Fig. \ref{figg2}a. In the same panel we display the first $9$ eigenfrequencies of this structure as a function of mode index $n$, whereas their corresponding electric field norms are shown in Fig. \ref{figg2}b. Note that the second EM mode, $n=2$, is resonant with the QEs, i.e., $\omega_2 = 1.4$ eV. The number of emitters in this system is $N=1.88\times10^4$, maintaining the same density as in the infinite nanowire case. Finally, a purely monochromatic pulse is chosen, i.e., $f(t) = 1$, which, as a result of the various loss mechanisms, will eventually lead the system into its steady state. We have calculated the steady-state wavefunction $\vert \psi\rangle$ both with and without dipole-dipole interaction $V_{ij}$, in order to have a more complete picture. In the former case, we account for disorder by performing a statistical average over $10^4$ realizations, each of them including a random inhomogeneous broadening in the energy of the QEs, $\epsilon_j \to \epsilon_j + \Delta_j$. The random broadening rate $\Delta_j \in [-\gamma_\phi,\gamma_\phi]$ is bound by the dephasing rate of J-aggregated molecules ($\gamma_\phi \sim 25$ meV) \cite{ValleauJCHEMPHYS2012}.

\begin{figure}
\centering
\includegraphics[width=\linewidth]{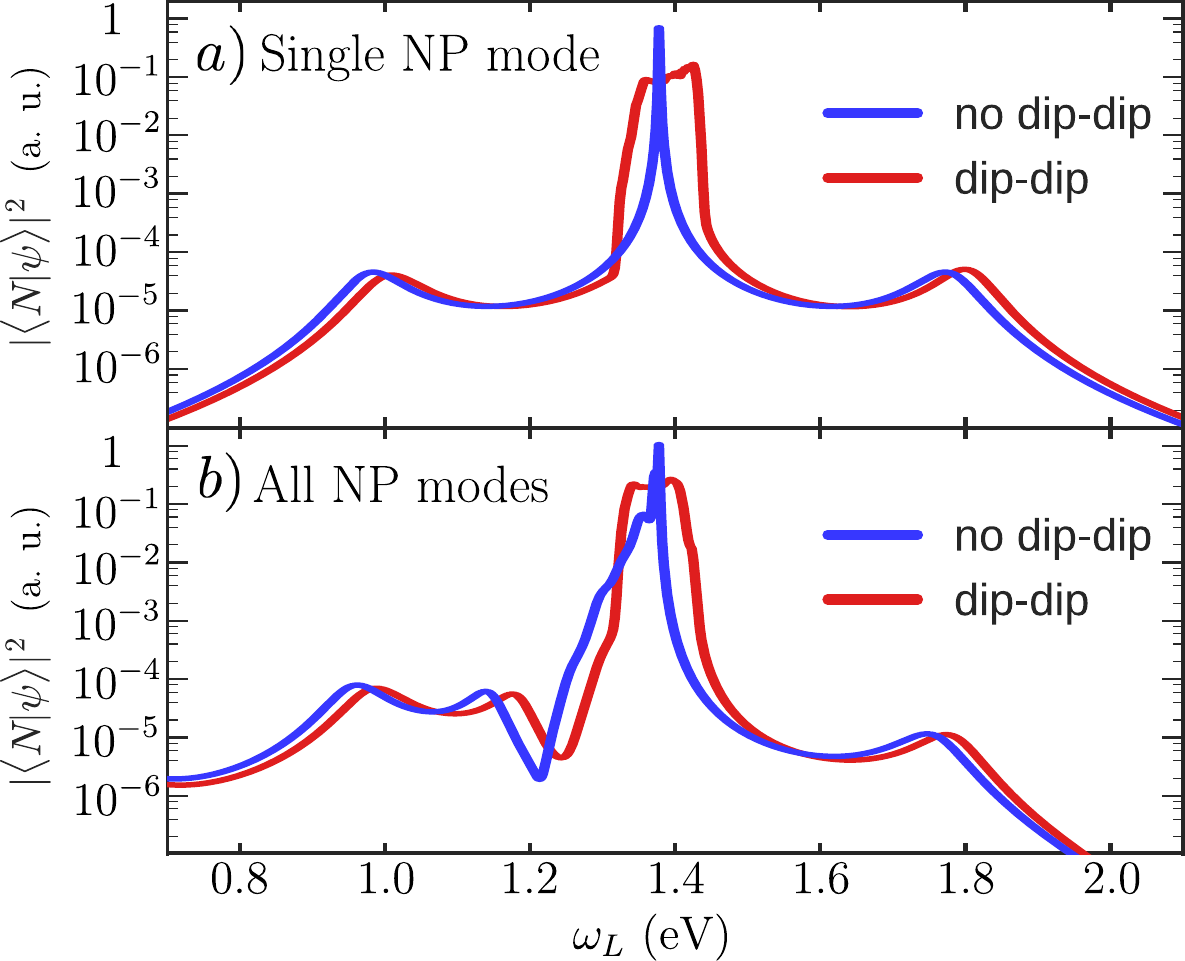}
\caption{(Color online) Steady-state population of the final emitters in the chain, as a function of the pump frequency $\omega_L$. a) Situation in which only the resonant photonic mode $n=2$ is taken into account. b) Same results when including all the modes. Red (Blue) lines show the case where nearest-neighbor dipole-dipole interaction is (not) included in the Hamiltonian.}\label{figg3}
\end{figure}

Let us consider first the case where only the resonant LSP mode $(n=2)$ is included in Eq. (\ref{H}). For this situation and neglecting dipole-dipole coupling between the QEs, we render in Fig. \ref{figg3}a (blue curve) the steady-state population of the QEs lying farthest from the pump region, $\vert \langle N \vert \psi \rangle\vert^2$, as a function of the pump frequency $\omega_L$. Notice that similar plots displaying a clear three-peak spectrum are obtained for the populations of every emitter in the ensemble. Therefore, the three maxima in the figure correspond to extended states, where the population is largely delocalized across the system. The two peaks at higher and lower energies are associated with the wavefunctions of the two polaritons, which inherit the delocalized character of the photonic excitations thanks to their hybrid nature. However, the emergence of a peak located at the frequency of the dark states implies that the population of these modes also extends over the whole system. Notably, this population is several orders of magnitude larger than those of the two polaritons.  This is in sharp contrast with the results obtained for an infinite NW, in which the wavepacket of the dark states was localized just around the pump region. 
 
A peak in the population spectrum related to the dark states remains when the dipole-dipole interaction is taken into account, as shown by the red curve in Fig. \ref{figg3}a. As this peak in the population is a fingerprint of the delocalized character of the associated wavefunction, we can safely conclude that interactions within the dark subspace do not destroy the extended nature of the dark states in this confined EM system. Moreover, the delocalized character also persists when several EM modes supported by the NP are included in the Hamiltonian, as demonstrated in Fig. \ref{figg3}b. In this case, the additional plasmon modes form new polariton states that appear as new peaks in the population. The majority of these new polaritons, namely those largely detuned with respect to the energy of the QEs, have a very small photonic component and thus form a band at around $1.16$ eV. Although the population spectrum is modified due to the presence of several EM modes in the system, the larger population associated with the dark modes as compared to those of the polaritons is maintained when the full spectrum of the EM environment is taken into account. This implies that our main finding, namely that dark states can inherit the delocalized character of the polaritons despite the fact they do not directly interact with the photonic modes, is very robust against both dipole-dipole interactions between the QEs and light fields that support several discrete EM modes.

To provide an analytical foundation for our main finding, we now elaborate a simple model that is able to capture the basic ingredients of the interaction of an ensemble of QEs with a photonic structure that displays a discrete EM spectrum. In this model we neglect dipole-dipole coupling and only consider a single EM mode since, as shown in Fig.  \ref{figg3}, these two effects play a minor role in the physical phenomenon under study. We also assume that, as in the numerical calculations presented above, the EM mode is resonant with the excitations within the QEs.  In this simple case, the eigenstates of the unperturbed Hamiltonian $H_0$ are formed by the $(N-1)$ dark states $\vert \text{D}\rangle$, and the upper and lower polariton $\vert \text{UP}\rangle$ and $\vert \text{LP}\rangle$, respectively. The former have the same energy as the bare QEs, $\epsilon_j \equiv \epsilon_0 -i\gamma_0/2$, where we have explicitly separated the real part from the loss rate of the QEs, $\gamma_0$. On the other hand, within the CSC regime ($\sqrt{N}g \gg \gamma_0,\gamma_m$), the energies of the states $\vert \text{UP}\rangle$ and $\vert \text{LP}\rangle$ are given by $\epsilon_\pm \approx \epsilon_0 -i (\gamma_0+\gamma_m)/4 \pm \sqrt{N}g$, where $\gamma_m$ is the loss rate of the light mode, and $g$ is the coupling rate of such mode to each of the QEs, which is assumed to be equal for all of them. The whole set of eigenstates of $H_0$ forms a complete set, i.e.,
\begin{equation}\label{closure}
\vert \text{UP}\rangle\langle\text{UP}\vert+\vert \text{LP}\rangle\langle\text{LP}\vert+\sum_D \vert \text{D}\rangle\langle\text{D}\vert=1.
\end{equation}

To analyze the dynamics, we start with the general expression for the system wavefunction $\vert \psi(t)\rangle$ that, to first order in the perturbative parameter $\Omega_p$, reads \cite{suppl}

\begin{equation}\label{psisolved}
\vert \psi(t)\rangle = \vert 0\rangle -i e^{-iH_0 t}\int_0^t dt' e^{iH_0 t'} V(t')\vert 0\rangle.
\end{equation}
By introducing the closure relation, Eq. \ref{closure}, into Eq. \ref{psisolved}, we can calculate the population probability amplitude for a generic emitter $j$, i.e., $p_j = \langle j \vert \psi(t)\rangle$, where $\vert j \rangle \equiv \sigma_j^\dagger \vert 0 \rangle$.  In the steady state, this magnitude is given by
\begin{widetext}
\begin{equation}\label{full}
p_j = -\frac{\Omega_p}{2}e^{-i\omega_L t}\bigg(\frac{\langle j \vert \text{UP}\rangle\langle\text{UP}\vert 1\rangle}{\epsilon_0-\omega_L + \sqrt{N}g-i(\gamma_0+\gamma_m)/4}+ \frac{\langle j \vert \text{LP}\rangle\langle\text{LP}\vert 1\rangle}{\epsilon_0-\omega_L-\sqrt{N}g-i(\gamma_0+\gamma_m)/4}+\sum_{\text{D}}\frac{\langle j \vert \text{D}\rangle\langle\text{D}\vert 1\rangle}{\epsilon_0-\omega_L-i\gamma_0/2}\bigg).
\end{equation} 
\end{widetext}

According to Eq. \ref{full}, the population $\vert p_j\vert^2$ will display three well-separated Lorentzian peaks centered at $\epsilon_0$ and $\epsilon_0\pm\sqrt{N} g$, respectively. Therefore,  Eq. \ref{full} is able to account for the numerical results as displayed in Fig. \ref{figg3}a. It is also straightforward to calculate the ratio between the population peak height associated with the dark modes and those associated with each of the two polaritons; this ratio is proportional to  $(1+\gamma_m/\gamma_0)^2$. In the case under study, loss associated with the predominant EM mode of the NP ($\gamma_m \approx 100$ meV, see Fig. \ref{figg2}a) is much larger than the loss rate of the QEs ($\gamma_0 \approx 1$ meV).  This explains why the population peak of the dark modes in Fig. \ref{figg3}a is four orders of magnitude higher than the heights of the polariton peaks. It is interesting to note that in the opposite limit, $\gamma_0>>\gamma_m$, our analytical formula predicts similar heights for the three population peaks. 

Finally, we can also understand the process of dark-state delocalization by first considering $\omega_L \approx \epsilon_0$ and introducing also the closure relation Eq. \ref{closure} into Eq. \ref{full}, obtaining

\begin{equation}\label{pjfinal}
p_j \big\vert_{(\omega_L \approx \epsilon_0)} \propto\frac{\langle j \vert \text{UP}\rangle\langle\text{UP}\vert 1\rangle+\langle j \vert \text{LP}\rangle\langle\text{LP}\vert 1\rangle}{\epsilon_0-\omega_L-i\gamma_0/2}.
\end{equation}
This expression shows that dark-state population can be expressed as a function of the two polaritons only. Since both these polaritons are spatially extended, dark states are therefore constrained to display the same delocalized behavior. Note that this is not a property of any particular dark state but of the dark subspace as a whole. In other words, by strongly coupling the QEs to a discrete electromagnetic mode, one extended state is removed from the QEs Hilbert space. This leaves an imprint on the remaining dark subspace, which hence inherits the delocalized character of the polaritons. Importantly, the dark states only acquires the delocalized nature of the polaritons but not their associated losses. As the dark modes do not couple with the EM modes, their losses are only governed by the loss rate of the QEs. When these loss rates are smaller than the radiative losses of the EM modes, dark modes become more efficient in transferring excitations across the system than polaritons. 

To conclude, in the collective strong coupling regime of an electromagnetic field to an ensemble of emitters, not only the polaritons but also the dark states can feature a delocalized behavior across the system. This unforeseen result, given the fact that dark states are uncoupled to light, is of a very general nature requiring only the discrete character of the relevant electromagnetic spectrum. While dark states delocalization is inherited from the corresponding polaritonic behavior, losses are not.  This is very advantageous when the population decay is dominated by photon absorption. Resonant structures with low to moderate quality factors can thus find a broad range of applications thanks to this different perspective on the properties of strongly coupled systems. 

This work has been funded by the European Research Council (ERC-2011-AdG Proposal No. 290981), the Spanish MECD (FPU13/01225 fellowship), the Spanish MINECO (MAT2014-53432-C5-5-R grant), and by the European Union Seventh Framework Programme under grant agreement FP7-PEOPLE-2013-CIG-618229.

\bibliographystyle{apsrev4-1}
\bibliography{bibliography}

\pagebreak

\renewcommand{\theequation}{A.\arabic{equation}}

\onecolumngrid
\begin{center}
\textbf{\large Supplemental Material}
\end{center}
\setcounter{equation}{0}
\setcounter{figure}{0}
\setcounter{table}{0}
\setcounter{page}{1}
\makeatletter

\vspace{0.9cm}

\twocolumngrid

\section{Calculation of the plasmon eigenmodes.}

In order to calculate the eigenmodes supported by the two plasmonic structures described in the main text, we use a Drude-Lorentz model for the silver permittivity,
\begin{equation}\label{DrudeLorentz}
\epsilon(\omega) = \epsilon_\infty -\frac{\omega_p^2}{\omega(\omega + i\gamma_D)}-\Delta\frac{\Omega_P^2}{\omega^2 - \Omega_P^2 +i\omega\Gamma_P},
\end{equation}
where the parameters $\epsilon_\infty\!=\!3.91$, $\omega_p\!=\!8.833\ $eV, $\gamma_D\!=\!0.0553\ $eV, $\Delta\!=\!0.76$, $\Omega_P\!=\!4.522\ $eV, and $\Gamma_P\!=\!8.12\ $eV are taken from Ref.~\cite{zhiming2008}.  For simplicity, we solve the eigenvalue Maxwell Equations for the lowest energy branch $m=0$. Here, $m$ represents the azimuthal number labeling any solution of an eigenmode equation in cylindrical coordinates $(\rho,z,\phi)$. Such value determines the azimuthal dependence of the solutions (in this case, the electric and magnetic fields) through $E(\rho,z,\phi) = E(\rho,z) e^{im\phi}$.

In the case of the nanowire, the dispersion $\omega(\kappa_\parallel)$ is extracted from the standard trascendental mode equation \cite{jackson1999classical}. Although usually the solution is expressed in terms of a real frequency and an complex parallel momentum, here we solve for a real  wavevector $k_\parallel$ and complex $\omega$ in order to comply with the picture introduced by the Hamiltonian (1) in the main text. With this convention, the calculation of the propagation length can still be carried out by using the group velocity $v_g = \partial(\text{Re}[\omega])/\partial k$. Indeed, we can employ the substitution $e^{-x/2L_p} = e^{-x \text{Im}[k]} \rightarrow e^{-x \text{Im}[\omega]/vg}$ to extract the plasmon propagation length as $L_p = (\partial(\text{Re}[\omega])/\partial k)/2\text{Im}[\omega]$. Finally, once the dispersion relation is determined, the nanowire modes can be extracted by using the analytical solution in terms of Bessel and Hankel functions \cite{jackson1999classical}.

\begin{figure}
	\centering
	\includegraphics[width=\linewidth]{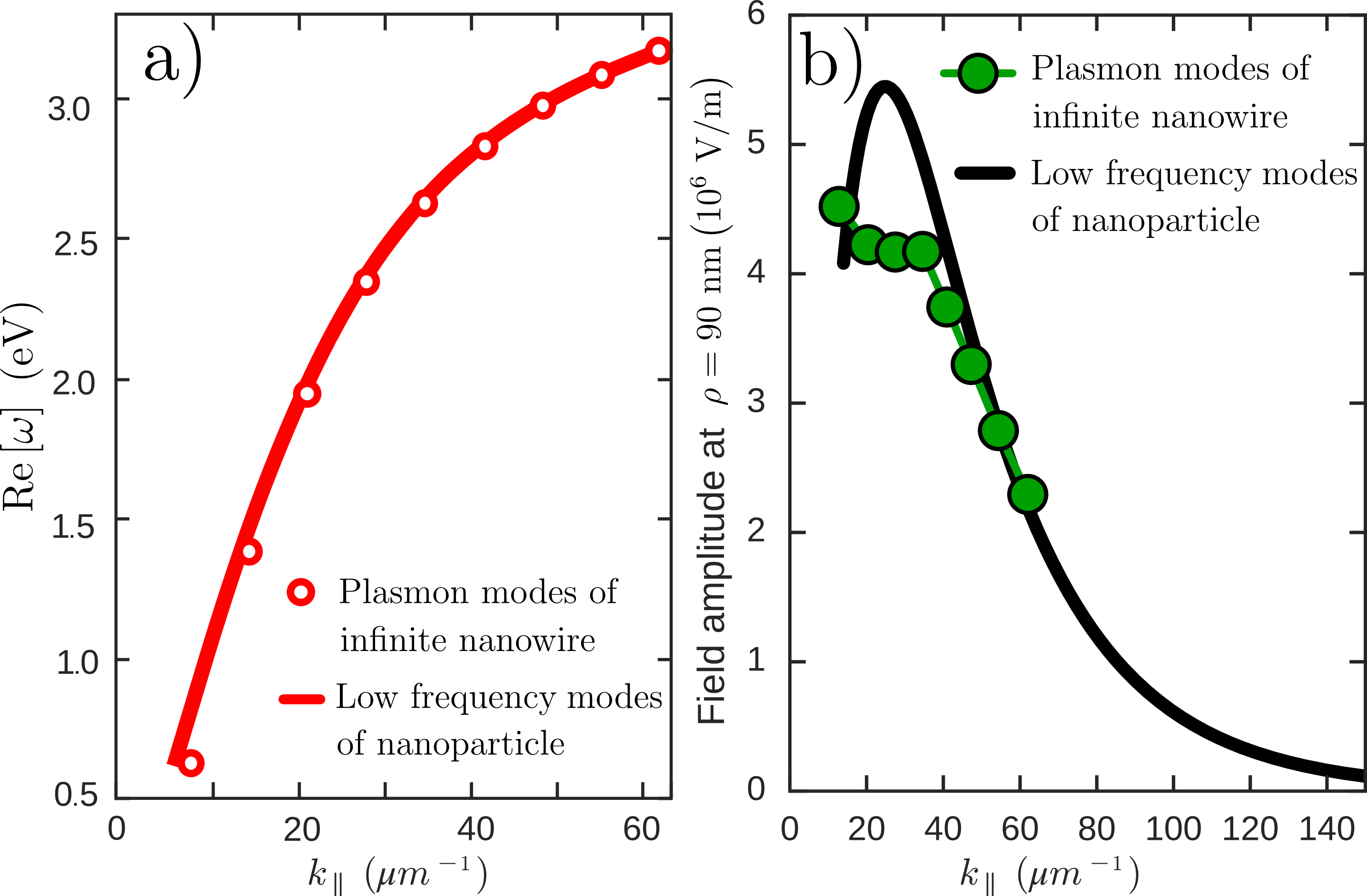}
	\caption{(color online) Comparison of the eigenmodes of the nanoparticle and the infinite nanowire. a) dispersion relation. b) radial field amplitude $35$ nm above the cylindrical surface.}\label{fig}
\end{figure}

For the nanoparticle, on the other hand, the lower energy eigenmodes have been calculated numerically. For large frequencies, however, the eigenenergies grow closer to each other, and the numerical computation becomes increasingly challenging. We avoid this problem by noting that, in the high energy limit, the relevant modal properties of the localized nanoparticle can be extracted from the modes of the infinite nanowire. This is demonstrated in Fig. \ref{fig}a, which compares the dispersion relation of both systems. For the nanoparticle, the allowed values of $k_\parallel$ have been extracted from the standing wave patterns of the electric field in our calculated low frequency eigenmodes (see e.g. Fig 2a of the main text). Since both curves converge to each other in the high energy limit, we can extrapolate the first LSP modes for large frequencies by using the analytical solutions for the nanowire.

The calculation of the $n-$th mode of the nanoparticle thus starts by obtaining  its parallel wavevector $k_\parallel$, by a direct extrapolation of the corresponding low energy values. This is straightforward, as $k_\parallel$ depends linearly on $n$. After, we introduce such wavevector in the dispersion relation for the infinite nanowire, in order to obtain the eigenfrequency of the mode, Re$\left[\omega_n\right]$. The imaginary part of $\omega_n$ is not accurately described by the NW solution, since the field concentrated around the two semispherical caps makes the nanoparticle modes much lossier than the NW plasmons. However, we have checked that the particular values of Im$\left[\omega_n\right] \; (n\gg 1)$ do not affect the physics of our system, as high-energy modes are poorly coupled to the QEs and their decay rates play a minor role in the dynamics.

Finally, it is necessary to find the electric field in a cylindrical layer surrounding the nanoparticle, i.e. the surface where the QEs will lie. Assuming that the nanoparticle is centered at the origin, the following extrapolation is an excellent approximation for $-L/2 \le z \le L/2$ and $n > 2$,
\begin{equation}
E \approx \bigg\lbrace\begin{array}{lcr}
E_0(n) \cos(k_\parallel (z-L/2)) & \text{for} & n \text{ odd}\\
E_0(n) \sin(k_\parallel (z-L/2)) & \text{for} & n \text{ even},
\end{array}
\end{equation}
where $E_0(n)$ is a field amplitude, dependent on the radial position $\rho$ of the considered surface. When this coordinate is equal to the separation of the QEs to the metallic surface ($\rho = r_0+35$ nm), the field amplitude $E_0(n)$ is very well approximated by the solutions of the infinite NW, as we show in Fig. \ref{fig}b. Note, finally, that the field intensity, and consequently the coupling rate $g_{jn}$, decays exponentially for large energies, and therefore we can reproduce the effect of all the nanoparticle eigenmodes with a finite number of modes in the Hamiltonian.

\section{Quantization of the EM fields.}

In the main text, the electromagnetic fields of the plasmon eigenmodes are calculated by solving the classical Maxwell equations. As in any eigenvalue problem, however, the obtained classical electric and magnetic fields  $\lbrace\mathbf{E}_{cl},\mathbf{H}_{cl}\rbrace$ are multiplied by an arbitrary normalization constant. Through the quantization procedure we fix such constant in order to introduce the electromagnetic fields in the quantum Hamiltonian (Eq. 1 in the main text).
We start by assuming the following general shape for the quantum field operators of each mode, $\hat{\mathbf{E}}$ and $\hat{\mathbf{H}}$,
\begin{equation}\label{qfields}
	\left(\begin{array}{c}
		\hat{\mathbf{E}}\\
		\hat{\mathbf{H}}
	\end{array}\right)= C \left(\begin{array}{c}
	\mathbf{E}_{cl}\\
	\mathbf{H}_{cl}
\end{array}\right) a + C^*\left(\begin{array}{c}
\mathbf{E}_{cl}^*\\
\mathbf{H}_{cl}^*
\end{array}\right) a^\dagger.
\end{equation}
Here, $a^\dagger$ and $a$ are the mode creation and annihilation operators, respectively, and $C$ is the normalization constant to determine. Next, we need to define the classical electromagnetic energy $U$. This energy is well approximated by \cite{maier2007plasmonics}
\begin{equation}\label{U}
	U \approx\!\! \int \!dV\!\! \left( \frac{\epsilon_0}{2} \frac{d\left(\omega\varepsilon_j(\omega)\right)}{d\omega}\bigg\vert_{\omega_\alpha} \!\!\!\!\mathbf{E}_{cl}\cdot \mathbf{E}^*_{cl} + \frac{\mu_0}{2} \mathbf{H}_{cl}\cdot \mathbf{H}^*_{cl} \right)\!,
\end{equation}
where $\varepsilon_j$ is the permittivity of each medium, and $\omega_\alpha$ is the frequency of the eigenmode. Note that the above expression assumes the losses in the metal to be low, i.e. Im$\left[\varepsilon_m\right]\ll$ Re$\left[\varepsilon_m\right]$, a very accurate approximation for low frequency modes. Although for high energy modes Eq. \ref{U} becomes less accurate, this does not affect the system dynamics since such modes are poorly coupled to the QEs and therefore play a minor role.

The next step in the quantization procedure consists in applying Bohr's correspondence principle to the classical energy and the quantum Hamiltonian $\hat{H}$, i.e. we perform the substitution $\lbrace U,\mathbf{E}_{cl}, \mathbf{E}^*_{cl},\mathbf{H}_{cl}, \mathbf{H}^*_{cl} \rbrace \rightarrow \lbrace \hat{H},\hat{\mathbf{E}}, \hat{\mathbf{E}}^\dagger,\hat{\mathbf{H}},\hat{\mathbf{H}}^\dagger \rbrace$
in Eq. \ref{U}. After expanding, we obtain
\begin{equation}
	\hat{H} = 2U \vert C \vert^2\left(a^\dagger a + \frac{1}{2}\right),
\end{equation}
where we have dropped terms proportional to $a^2$ and $\left(a^\dagger\right)^2$, whose contribution is negligible in the low loss limit. 
The above Hamiltonian has the usual harmonic oscillator form, where we can identify the normalization constant as
\begin{equation}
	\hbar \text{Re}\left[\omega_{\alpha}\right] = 2U\vert C \vert^2 \;\;\;\longrightarrow \;\;\; C = \sqrt{\frac{\hbar \text{Re}\left[\omega_{\alpha}\right]}{2U}}.
\end{equation}
thus, in order to determine $C$ from the classical modes we only need to calculate the classical energy $U$ through a volume integral. In the infinite nanowire, such integral is taken along a whole periodic unit cell.

\section{Calculation of the system wavefunction.}

For the numerical calculation of the wavefunction in both nanostructures, we use a reduced Hamiltonian since the large number of QEs greatly increases the computation time. In order to do so, we note that the QEs are regularly distributed in identical rings along the longitudinal coordinate of the cylinder, $z$. As we work in the weak driving regime $\Omega_p\ll 1$ and the initial state is the vacuum, the system will never abandon the single-excitation subspace, and we can therefore replace the ensemble of QEs in the Hamiltonian by a single chain of sites along the $z$ axis, each site describing a whole ring of QEs. In this picture, the original Hamiltonian must be expressed in terms of the operators
\begin{equation}
\tilde{\sigma}_j = \frac{1}{\sqrt{n_R}}\sum_{l\in \text{ ring }j}^{n_R}\sigma_{(l)j},
\end{equation}
which describe a collective excitation in the ring $j$ with the same angular distribution as the plasmon field, i.e. $m=0$. In the equation above, $n_R$ is the total number of emitters in the ring, and $\sigma_{(m)j}$ describes the fermionic annihilation operator for the emitter $m$ inside the ring $j$. Let us express the Hamiltonian $H_0$ as a function of these collective operators. First, the energy of the bare QEs does not change since
\begin{equation}
\left(\sum_j \epsilon_j \sigma_j^\dagger \sigma_j \right)\tilde{\sigma}_{j'}^\dagger\vert0\rangle = \epsilon_{j'} \tilde{\sigma}_{j'}^\dagger\vert0\rangle,
\end{equation}
where we assume all the QEs belonging to a given ring are identical. The above expression allows us to write the corresponding contribution to the Hamiltonian in terms of a sum not over all the emitters but over the different rings, i.e. 
\begin{equation}
\sum_j \epsilon_j \sigma_j^\dagger \sigma_j  = \sum_r \epsilon_r \tilde{\sigma}_r^\dagger \tilde{\sigma}_r
\end{equation}
where we explicitly use $r$ to describe an index running along all the rings in the ensemble. 

The interaction terms in the Hamiltonian, on the other hand, are modified. First, noting that the coupling $g_{j\alpha}$ is the same for all the QEs in a given ring, the term in $H_0$ describing the light-emitter coupling can be expressed as
\begin{widetext}
\begin{equation}
\sum_{j\alpha} g_{j\alpha} \sigma_j^\dagger a_\alpha \equiv \sum_{\text{ring }j,\alpha} g_{j\alpha} a_\alpha \left(\sum_{m\in \text{ ring }j}^{n_R}\sigma_{(m)j}^\dagger\right) = \sqrt{n_R} \sum_{ \text{ring }j,\alpha} g_{j\alpha} \tilde{\sigma}_j^\dagger a_\alpha = \sum_{r,\alpha} \left( \sqrt{n_R}g_{r\alpha} \right)\tilde{\sigma}_r^\dagger a_\alpha,
\end{equation}
\end{widetext}
i.e. this contribution to the Hamiltonian keeps its original form, with a new coupling intensity $\sqrt{n_R}$ times larger than the original. Finally, the total dipole-dipole interaction can also be expressed as a sum over rings as
\begin{equation}
H_{dd} = \sum_{\text{ring a}} \sum_{\text{ring b}} \sum_{m\in \text{ ring }a}^{n_R} \sum_{n\in \text{ ring }b}^{n_R}\sigma_{(m)a}^\dagger \sigma_{(n)b} V_{ab}^{mn}
\end{equation}
The dipole-dipole interaction between rings, $\tilde{V}_{ij}$, is thus defined through the following overlap,
\begin{equation}
\tilde{V}_{ij} = \langle 0 \vert \tilde{\sigma}_i H_{dd} \tilde{\sigma}_j^\dagger \vert 0 \rangle = \frac{1}{n_R}  \sum_{m\in \text{ ring }i}^{n_R} \sum_{n\in \text{ ring }j}^{n_R} V_{ij}^{mn},
\end{equation}
where the usual anticommutation relations have been used for the fermionic operators of the QEs. Since our system is axially symmetric, each sum in $i$ gives exactly the same result, and hence we can express the ring-ring interaction as
\begin{equation}
\tilde{V}_{ij} =  \sum_{n\in \text{ ring }j}^{n_R} V_{ij}^{0n} \equiv \sum_{n\in \text{ ring }j} V_{ij}^{n}.
\end{equation}
According to the above formula, the dipole-dipole interaction between rings $i$ and $j$ is the sum of the dipole-dipole coupling between \textit{one} emitter of ring $i$ and each of the QEs inside ring $j$. The final expression for the Hamiltonian $H_{dd}$ is therefore
\begin{equation}
H_{dd} = \sum_{r} \sum_{r'} \tilde{V}_{rr'} \tilde{\sigma}_r^\dagger \tilde{\sigma}_{r'} + H.c.
\end{equation}
By performing the substitutions above we recover a Hamiltonian with the same shape as the original one, $H_0$, in which both dipole-dipole interaction and QE-emitter couplings are modified such that every site in the Hamiltonian accounts for a whole ring of QEs.

For the analytical calculation of the wavefunction in the nanoparticle case, we start by explicitly extracting the small parameter $\Omega_p$ from the pump Hamiltonian,
\begin{equation}
V(t) = \Omega_p V_p(t),
\end{equation}
where the time-dependent pump is given by
  \begin{equation}\label{Vp}
  V_p(t) = \cos(\omega_L t)\left(\sigma_1^\dagger+\sigma_1\right).
  \end{equation}
We proceed by expanding the system wavefunction to first order in the weak pump intensity $\Omega_p$,
\begin{equation}\label{psi}
\vert \psi(t)\rangle = \vert \psi_0(t)\rangle + \Omega_p \vert \psi_1(t)\rangle + \mathcal{O}(\Omega_p^2),
\end{equation}
where the first term in the expansion represents the evolution of the initial state, $\vert \psi(0)\rangle$, by the unperturbed Hamiltonian,
\begin{equation}
\vert \psi_0(t) \rangle = e^{-iH_0 t}\vert \psi(0)\rangle.
\end{equation}
The next step is to apply the time-dependent Schr\"odinger equation to the state in Eq. \ref{psi}. By keeping only the linear terms in $\Omega_p$, the equation reduces to
\begin{equation}
i\left(\partial/\partial_t\right)\vert \psi_1(t) \rangle = V_p \vert \psi_0(t) \rangle + H_0 \vert \psi_1(t) \rangle.
\end{equation}
It is possible to eliminate the second term above by expressing the equation in the interaction picture. Then, we can formally integrate both sides of the equality and transform back into the Schr\"odinger picture, obtaining
\begin{equation}
\vert \psi_1(t) \rangle = -i e^{-iH_0 t}\int_0^t dt' e^{iH_0 t'} V_p(t')\vert \psi_0(t)\rangle.
\end{equation}
Finally, we assume the system is initially in the ground state, $\vert \psi(0)\rangle = \vert 0 \rangle$, which by definition has zero energy. Then, the final expression for the wavefunction takes the simple form
\begin{equation}\label{psisolved}
\vert \psi(t)\rangle = \vert 0\rangle -i\Omega_p e^{-iH_0 t}\int_0^t dt' e^{iH_0 t'} V_p(t')\vert 0\rangle,
\end{equation}
as seen in the main text.

Once the wavefunction has been calculated, it is straightforward to determine the probability amplitude $p_j = \langle j \vert \psi\rangle$,
\begin{equation}
p_j = -i\Omega_p \langle j \vert e^{-iH_0 t}\int_0^t dt' e^{iH_0 t'} \cos(\omega_L t')\vert 1\rangle,
\end{equation}
where we have substituted the expression for $V_p(t)$ (Eq. \ref{Vp}). We now introduce the closure relation $\sum_{\epsilon}\vert\epsilon\rangle\langle\epsilon\vert = 1$, where $\vert \epsilon\rangle$ are the eigenstates of the unperturbed Hamiltonian $H_0$. After a simple integration, we obtain
\begin{equation}\label{pijovs}
	\begin{split}
p_j =& -i\frac{\Omega_p}{2} \sum_{\epsilon} \langle j \vert\epsilon\rangle\langle\epsilon \vert 1\rangle\times\\ &\left(\frac{e^{i\omega_Lt}-e^{-i\epsilon t}}{\epsilon+\omega_L} + \frac{e^{-i\omega_L t}-e^{-i\epsilon t}}{\epsilon-\omega_L}\right) .
\end{split}
\end{equation}
In the expression above, we can neglect the off-resonant terms $\propto (\epsilon + \omega_L)^{-1}$. Additionally, since all the eigenstates of the Hamiltonian suffer from losses, in the steady state limit $t\to\infty$ the exponentials $e^{-i\epsilon t}$ vanish. Therefore, the steady-state amplitude can be expressed as
\begin{equation}\label{pijovs2}
p_j \approx -i\frac{\Omega_p}{2} e^{-i\omega_L t}\sum_{\epsilon} \frac{\langle j \vert\epsilon\rangle\langle\epsilon \vert 1\rangle}{\epsilon-\omega_L} ,
\end{equation}
 an equivalent expression to  Eq.  5 in the main text.

\end{document}